# Coded Single-Tone Signaling and Its Application to Resource Coordination and Interference Management in Femtocell Networks

C. Yang, C. Jiang, J. Wang

*Abstract*—Resource coordination and interference management is the key to achieving the benefits of femtocell networks. Over-the-air signaling is one of the most effective means for distributed dynamic resource coordination and interference management. However, the design of this type of signal is challenging. In this paper, we address the challenges and propose an effective solution, referred to as coded single-tone signaling (STS). The proposed coded STS scheme possesses certain highly desirable properties, such as no dedicated resource requirement (no overhead), no near-and-far effect, no inter-signal interference (no multi-user interference), low peak-to-average power ratio (deep coverage). In addition, the proposed coded STS can fully exploit frequency diversity and provides a means for high quality wideband channel estimation. The coded STS design is demonstrated through a concrete numerical example. Performance of the proposed coded STS and its effect on cochannel traffic channels are evaluated through simulations.

## I. Introduction

Femtocell networks have been gaining a significant amount of interest for improving coverage and/or increasing capacity by higher spatial reuse via cell splitting in modern wireless communication networks [1]-[3] . Due to the unplanned and dense deployment nature and the restricted access property (a user may be denied for access to the nearest base station and forced to obtain service from a farther base station thereby the received power from the interferer can be much stronger than that from its serving base station) of femtocell networks, interference can be so severe such that if not efficiently managed the benefits of femtocell networks may not be gained [4]-[10].

Tighter coordination among neighboring base stations to ensure QoS and fairness across cells in femtocell networks thus becomes even more crucial than that in the conventional macrocell network. A straight forward solution is resource partitioning among interfering base stations [11]. This static partitioning scheme makes sense for low bandwidth user-dedicated control channels but is obviously resource-inefficient for data transmission, especially in a dense deployment environment where an overloaded base station can only utilize a small portion of the resources whereas most of the resources could be held up by other idling base stations. Dynamic resource allocation hence makes more sense for data transmission in terms of resource usage [12][13]. In this approach, resources are negotiated among interfering base stations or scheduled by a central controller through signaling over the backhaul on an on-demand and maximization of network utility (e.g., fairness, data rates and delay of QoS flows) basis. However, unlike in macrocell networks where base stations are connected via exclusive/leased backhaul (e.g., the X2 interface in LTE), direct connection among femto base stations is impractical. A femtocell is typically connected to the cellular core network via a third party IP backhaul [14]. The delay can be significant and may vary from 10s of msec to 100s of msec. The large and yet unpredictable delay makes signaling through femtocell backhaul unsuitable for dynamic resource coordination and interference management in the case of bursty and delay sensitive applications. Over-the-air signaling is thus a better choice for this type of applications where the coordination message is sent to the interfering base stations over the air by the user since the base stations that hear the message are the strongest interferers to this user.

However, over-the-air signaling poses its own unique challenges. First, over-the-air signaling does not come free. It requires considerable precious wireless resource for the reliable transmission of the coordination message; Second, it requires deep coverage, since the target receivers for the coordination message are the *neighboring base stations* (not the serving base stations); Third, due to the broadcast nature of over-the-air signaling, conventional random access signaling [15]-[18] suffers from interference among signals sent by different users, and the well-known near-and-far effect [19]-[21]. Since the target receivers are the neighboring base stations, the near-and-far effect for this type of application is particularly damaging. In the existing cellular systems, CDMA is used for random access, such as the random access signals used in uplink random access channels (RACHs) [22]-[24]. This signal is used by users to request channel resources from its *serving* base station. To minimize the near-and-far effect, the transmit power must be carefully managed. The user first determines the minimum transmit power according to the received signal strength of the downlink pilot signal and gradually increases its transmit power at each failed attempt in order to avoid blanking out other users' access signals. This scheme clearly does not apply to the current application since the recipients of the coordination message are the neighboring base stations. Maximum power is typically required to transmit the coordination message to ensure sufficient coverage which inevitably blocks its serving base station from hearing the signals from the users of other cells.

In this paper, we propose a signaling scheme, referred to as coded single-tone signaling, that is particularly efficient for femtocell coordination application. We will focus on downlink resource and interference management in which the coordination messages are transmitted on the uplink by the user to the interfering base stations since the base stations that hear the message are the strongest interferers.

The rest of this paper is organized as follows. Section II gives a detailed description of the coded single-tone signaling scheme. Section III analyzes the proposed coded STS tone detection and decoding analysis. Section IV provides a concrete design example of the coded STS and the numerical results. Section V concludes this paper.

## II. CODED SINGLE-TONE SIGNALING

In the proposed single-tone signaling (STS), a large fraction, if not all, of the energy in an OFDM symbol is transmitted on a single OFDM subcarrier. No energy is transmitted on any other subcarriers of the current OFDM symbol. No information is modulated onto the energized subcarrier (i.e. amplitude and/or phase). It is the *location* (subcarrier index) of the energized tone that contains information. That is, which subcarrier of this OFDM symbol is energized depends on the content of the message that STS carries. The message for resource coordination and interference management information is denoted as *m*, which is further represented by *K* information symbols, $\mathbf{u} = [u_1, u_2, ..., u_K]^T$, where $0 \leq u_k \leq D-1$ for $1 \leq k \leq K$, or more precisely,

$$m(D) = u_K D^{K-1} + u_{K-1} D^{K-2} + ... + u_2 D + u_1 \qquad (1)$$

where $u_k$, $1 \leq k \leq K$, is the *index* of the subcarrier that the STS tone is transmitted on, the base $D \ (\leq S)$ is the total number of subcarriers of an OFDM symbol used for STS transmission, and $S$ is the total number of subcarriers in an OFDM symbol. Without loss of generality, we assume $D = S$ in this paper. We hence need *K* OFDM symbols to transmit the message *m*, as shown in Fig. 1.

Note that an STS signal is not to be confused with a frequency hopping signal. In frequency hopping, the tone positions are *predetermined* by a sequence known to both the transmitter and the receiver; the tone position therefore does not contain information and the information is modulated onto the amplitude and/or phase of the tone via QPSK or QAM. While for STS, the tone is not modulated with information. The information is embedded in the positions/subcarrier indices of the tones.

The choice of this type of signal has the following advantages. First of all, unlike the most commonly used CDMA signals for random access [15]-[23], the STS does not suffer from the near-and-far effect among STS transmissions from different users. This is because 1) If the STS signals from different users are transmitted on different subcarriers, they don't present interference to each other since all the OFDM subcarriers are orthogonal; 2) If some of the STS signal tones from different users happen to transmit on the same subcarrier, they simply add together just like multipaths (since the STS tones are not modulated therefore share the same waveform and are not distinctive at the receiver) and are absorbed by cyclic prefix[25]. This property is crucial for this particular application since the target receivers of the signal are the neighboring base stations rather than the serving base station. Second, an STS tone is a narrow-band constant-profile sinusoidal waveform, therefore has low peak-to-average power ratio (PAPR) allowing for a higher power amplifier setting and thereby deeper coverage [26][27]; Third, since the transmit energy is concentrated on one single subcarrier of an OFDM symbol, the STS tone is much stronger than a regular data tone therefore is easy to be detected even under strong interference environment. If such a strong tone that reaches a base station becomes hard to be detected, this base station will not likely cause significant interference to the sender, i.e., the sender (the mobile user) must be out of the interference range of the base station. As a result, STS transmission can be overlaid with other users' uplink data transmission. On the other hand, the interference to the uplink data traffic is also concentrated on a subcarrier of an OFDM symbol. This isolated interference can be most effectively removed by the data decoder, thereby causing minimal impact to the data decoding since the decoder is very effective in removing isolated erasures/errors. This means that an STS signal can be transmitted without designated system resource (Therefore, other users don't need to clear the subcarriers for STS). Hence no overhead is incurred. This is in contrast to the conventional random access signaling scheme where a continuous chunk of the total frequency band is set aside for random signaling by multiple users using, for example, PN or Z-C sequences based on CDMA technology [21][28].

Despite the advantages of STS, however, single tone signaling can also be prone to errors due to fading (falsely detected tones or missed detection of tones). Furthermore, even under perfect detection, i.e., no falsely detected tones and missed tones, simultaneous transmissions of STS signals from multiple users may confuse the receiver. For example, if more than one STS tones from more than one users are detected in one OFDM symbol, the receiver will not be able to distinguish the STS tones from different users (cf. Fig. 2) due to the fact that STS tones are not user-distinctive as earlier stated, causing ambiguity to the receiver. The possible combinations are $t^K$ where $t$ is the number of simultaneous transmitting users ($t = 2$ in Fig. 2). Hence it is necessary to encode the STS signal (non-binary) to obtain certain degrees of error protection and ambiguity prevention capability. Reed-Solomon codes are non-binary codes and achieve the largest possible code minimum distance for any linear code with the same encoder input and output block lengths (or maximum distance separable). Hence the Reed-Solomon code is a good fit for the current application. A Reed-Solomon code $(N, K)$ encodes the non-binary information symbols $\mathbf{u} = [u_1, u_2, ..., u_K]^T$ into a codeword $\mathbf{c} = [c_1, c_2, ..., c_N]^T$, where $0 \leq c_n \leq D-1$ for $1 \leq n \leq N$. The error-correction capability is $t = \left\lfloor \dfrac{N-K}{2} \right\rfloor$ and the erasure-correction capability is $\rho = N - K$, where $\lfloor x \rfloor$ denotes the maximum integer which does not exceed $x$.

Let $\alpha$ be a primitive number in Galois Field GF($D$), the $K$ information symbols $\mathbf{u} = [u_1, u_2, ..., u_K]^T$ are encoded into an $N$ coded symbols via the Galois Fourier Transform [29]

$$\mathbf{c}^T = \mathbf{Z} \begin{bmatrix} 0 & \mathbf{u}^T & 0 & \cdots & 0 \end{bmatrix}^T \tag{2}$$

where

$$\mathbf{Z} = \begin{bmatrix} 1 & 1 & \cdots & 1 \\ 1 & \alpha^{\frac{D-1}{N}} & \cdots & \alpha^{\frac{D-1}{N}(N-1)} \\ \vdots & \vdots & \ddots & \vdots \\ 1 & \alpha^{\frac{D-1}{N}(N-1)} & \cdots & \alpha^{\frac{D-1}{N}(N-1)(N-1)} \end{bmatrix} \tag{3}$$

The value of the coded symbol ($c_n$, $1 \leq n \leq N$) corresponds to the *index* of the sub-carrier on which energy is transmitted. That is, only one tone is energized per OFDM symbol, and the position of the tone is dependent on the value of the coded symbol. Either all or partial of the total energy in an OFDM symbol is transmitted on a single subcarrier depending on the desired coverage range. Fig. 3 shows an example of the transmission of the coded STS signal.

## III. STS Detection and Decoding

### A. STS Tone Detection Analysis

STS tone detection is the first step of coded STS detection and decoding. It is important to note that the STS tone by itself is not user-distinctive (cf. Section II). *The base station receiver should detect all these tones without user discrimination. It is the STS decoder's job to separate users' messages based on all the received STS tones (will be discussed in the next subsection).* This sub-section is devoted to the analysis of STS tone detection performance.

The received signal vector per tone at a receiver with $N_r$ receive antennas from $N_t$ transmit antennas can be described as follows:

$$\mathbf{y} = \begin{cases} \mathbf{H}\sqrt{\mathbf{P}} + \mathbf{n}, & \text{STS tone present} \\ \mathbf{n}, & \text{Otherwise} \end{cases} \quad (4)$$

where $\mathbf{y} = [y_1, \ldots, y_i, \ldots, y_{N_r}]^T$ denotes the received signal vector from $N_r$ receive antennas;

$$\mathbf{H} = \begin{bmatrix} h_{11} & \cdots & h_{1N_t} \\ \vdots & \ddots & \\ h_{N_r 1} & \cdots & h_{N_r N_t} \end{bmatrix} \quad (5)$$

denotes the channel gains between the $N_t$ transmit antennas and $N_r$ receive antennas, and $h_{ij}, 1 \leq i \leq N_r, 1 \leq j \leq N_t$ is distributed according to standard complex Gaussian distribution $CN(0,1)$; $\sqrt{\mathbf{P}} = \left[\sqrt{p_1}, \ldots, \sqrt{p_j}, \ldots, \sqrt{p_{N_t}}\right]^T$ and $p_j, 1 \leq j \leq N_t$ denotes the transmit power of transmit antenna $j$; $\mathbf{n} = \left[n_1, \ldots, n_i, \ldots, n_{N_r}\right]^T$ denotes the noise vector at receiver from $N_r$ receive antennas and $n_i, 1 \leq i \leq N_r$ can be modeled as a complex Gaussian variable distributed according to $CN(0, \sigma_i^2)$.

According to the above assumption, random variable $y_i$ is distributed according to

$CN(0, \sum_{j=1}^{N_t} p_j + \sigma_i^2)$, $1 \leq i \leq N_r$ when STS tone is present; and $y_i$ is distributed according to $CN(0, \sigma_i^2)$, $1 \leq i \leq N_r$ when STS tone is absent.

After combining the energy from multiple receive antennas, the detection variable (reflecting the combined energy detected per tone) is:

$$z = \sum_{i=1}^{N_r} |y_i|^2 \qquad (6)$$

According to the theory in [30], the error probability (the probability an STS tone is detected on a subcarrier where the STS tone is absent) can be expressed as:

$$P_{\text{error}} = P_0(z \geq x) = 1 - F_0(x) = \sum_{k=0}^{N_r - 1} \frac{1}{k!} \left( \frac{x}{\sigma_{k+1}^2} \right)^k \exp\left( \frac{-x}{\sigma_{k+1}^2} \right), \; x \geq 0 \qquad (7)$$

where $x$ is the detection threshold. By controlling the threshold $x$, we can control the probability of the false detection. Since the STS tones are much stronger than the user regular data tones, $x$ can be set relatively high to prevent data tones from being detected as STS tones. For the base stations, where the STS tones from the user are buried in the data tones, these base stations must be far away from the user who sent the STS (hence not significant interferers to this user). Therefore, failure to detect the STS by these far-away base stations is expected by design.

The erasure (missed detection) probability of an STS tone can be obtained as:

$$P_{\text{erasue}} = P_1(z < x) = F_1(x) = 1 - \sum_{k=0}^{N_r - 1} \frac{1}{k!} \left( \frac{x}{\sigma_{k+1}^2 + \sum_{j=1}^{N_t} p_j} \right)^k \exp\left( \frac{-x}{\sigma_{k+1}^2 + \sum_{j=1}^{N_t} p_j} \right), \; x \geq 0 \qquad (8)$$

When a subcarrier that contains STS tones from $N_{user}$ different users arrives at the same instant, the erasure probability of the STS tone can be obtained as:

$$P_{\text{erasue}}^{N_{user}} = 1 - \sum_{k=0}^{N_r - 1} \frac{1}{k!} \left( \frac{x}{\sigma_{k+1}^2 + N_{user} p} \right)^k \exp\left( \frac{-x}{\sigma_{k+1}^2 + N_{user} p} \right), \; x \geq 0 \qquad (9)$$

From (9), we can get $P_{\text{erasue}}^{N_{user}} < P_{\text{erasue}}^1$, for $N_{user} > 1$. This means that multiple users transmitting STS on the same subcarrier does not worsen but improves the detection performance. which is not surprising since STS tones share the same waveform and receive power gain can be obtained from different users. When STS tones on the same OFDM subcarrier from different users arrive with different delays, they behave just like multipath signals whose energies can be absorbed by the OFDM cyclic prefix, presenting no interference to each other.

### B. STS Decoding

After the detection of STS tones, the receiver obtains a set of subcarrier indices on which STS tones are detected on every OFDM symbol with certain false detection errors as well as missed STS tones. By applying, for example, maximum likelihood decoding, the receiver is able to recover the original information symbols.

In the presence of multiple STS signals, the STS signals from different users may overlay on top of each other causing potential ambiguity at a receiver (c.f., Fig. 2). Indeed, $d$ STS signals with code rate $(N, K)$ can coexist without causing decoding ambiguity as long as the following inequality

$$K \leq \left\lceil \frac{N}{d} \right\rceil \tag{10}$$

is satisfied. This important conclusion can be formally stated by the following proposition:

*Proposition: Assume $d$ ($d \leq D^K$) distinctive STS signals coded on GF(D) with code rate $(N, K)$ are simultaneously received on the same time and frequency resource. Under perfect tone detection, all d coded STS signals can be decoded to the original information symbols without ambiguity, if $K \leq \left\lceil \frac{N}{d} \right\rceil$ is satisfied.*

*Proof:* Consider $d$ $\left(d \leq D^K\right)$ distinctive STS signals coded with rate $(N, K)$ on *GF(D)* are simultaneously received from $N$ OFDM symbols, free of tone erasures and detection errors. Now arbitrarily

select *N* number of the detected tones, each from one of the *N* different OFDM symbols. We maintain that

1) There are at least $\left\lceil \dfrac{N}{d} \right\rceil$ STS tones out of the *N* selected tones coming from the same STS signal among the total number of *d* STS signals. This is the direct outcome from the pigeonhole principle.

2) For an STS signal with code rate of $(N, K)$, a minimum number of *K* STS tones is sufficient to distinguish one STS signal from another. This is sustained by the fact that Reed-Solomon codes are maximum distance separable.

We therefore conclude that if $\left\lceil \dfrac{N}{d} \right\rceil \geq K$, the *d* STS signals can be uniquely separated from each other without ambiguity. □

For example, $(16, 2)$ coded STS signals can allow up to 15 simultaneous STS signal transmissions from different users without causing confusion at a receiver.

However, the case of particular interest is *K*=1. When *K*=1, (10) holds for *any* value of *d* regardless the value of *N*. We therefore have the following important remark:

*Remark 1*: *Assume d ($d \leq D$) distinctive STS signals coded on GF(D) with code rate $(N,1)$, $\forall N \geq 1$ are simultaneously received on the same time and frequency resource. Under perfect tone detection, all d STS signals can be decoded to the original information symbols without ambiguity.*

Two users transmitting two *distinctive* STS signals do not collide with each other. For example, (8, 1) coded STS signals in theory can support up to *D* number of simultaneous STS signal transmissions. This is true whether the value of *N* equals to 8 or not.

However, this conclusion is only true under the assumption of ideal tone detection. In practical scenarios *when STS tone detection is not error-free* due to fading and noise. the value of *N* does affect the STS signal's capability of correcting tone detection errors,. The choice of *N* and the capability of STS tone error correction and erasure recovery have been discussed in Section II. Therefore, as long as the number of STS tone detection errors/erasures are within the capability of the coded STS signals governed by the value of $N$, multi-user ambiguity can be eliminated.

Another problem that can not be overlooked in the over-the-air signaling design for femtocell networks is the time and frequency synchronization. Time and frequency among femtocells may not be as perfectly synchronized as in macrocells. Since the STS signal is meant to be received by neighboring base stations, the STS has to be designed with time and frequency offset tolerance. The time offset among femtocells is easily absorbed by the cyclic prefix of the OFDM symbol [25] thereby is less of a concern. The design of STS with frequency offset immunity is more challenging since the information that an STS signal carries lies in the subcarrier index of the STS tones. If not appropriately designed, the STS tones may shift to neighboring subcarriers as a result of frequency offset. Again, we have to rely on coding to provide the frequency offset immunity.

Assume that the transmitted STS signal is

$$\mathbf{c} = \begin{bmatrix} c_1 & c_2 & \cdots & c_N \end{bmatrix}. \tag{11}$$

From (2), the inverse GFT of (11) is given by

$$\mathbf{Z}^{-1}\mathbf{c}^T = \begin{bmatrix} 0 & \mathbf{u}^T & 0 & \cdots & 0 \end{bmatrix}^T \tag{12}$$

with the first element equal to zero. Therefore, the inverse GFT of a valid code word always has a zero-valued first element *by design*.

However, the STS signal, transmitted by a user and received by a neighbor femtocell base station with frequency offset $\delta$, is

$$\mathbf{c}' = \begin{bmatrix} c_1 + \delta & c_2 + \delta & \cdots & c_N + \delta \end{bmatrix}. \tag{13}$$

The first element of the inverse GFT of (13) produces

$$\frac{1}{N}\sum_{n=1}^{N}(c_n + \delta) = \frac{1}{N}\sum_{n=1}^{N}c_n + \delta = \delta \tag{14}$$

which is non-zero (for $\delta \neq 0$). We therefore have the following remarks:

*Remark 2: An STS signal received with a frequency offset does not correspond to a valid STS signal.*

This property ensures that a receiver with a frequency offset to its transmitter will not erroneously map an STS signal to a valid (but wrong) message.

*Remark 3: The value of the first element of the inverse GFT of a received STS signal equals to the frequency offset between the sender and the receiver of the STS signal.*

This property enables the receiver to correct the frequency offset, if any, between the sender and the receiver. The frequency-offset STS signal can then be recovered. This property is hence particularly beneficial to femtocell networks where base stations installed by customers may not be perfectly frequency synchronized to each other.

## IV. NUMERICAL RESULTS

In this section, we provide a concrete coded STS design example for use in a femtocell network. We will adopt the LTE framework [31] and focus on the downlink for the following discussion. As earlier stated, the coded STS is used for resource coordination and interference management in femtocell networks. A user who is experiencing severe interference from one or more base stations, due to, e.g., entering the coverage of femtocell base stations with restricted access, broadcasts the resource coordination request message (RCRM) via coded STS to all neighboring base stations. The base stations that are able to decode the RCRM (therefore, major interferers) can simply clear the resource based on certain criterion or can coordinate power and spatial beams according to the received STS.

The information included in the RCRM in general can be the assigned radio resource identity by the serving base station, traffic priority indicator, the target SINR indicator, etc.

Specifically, the assigned radio resource identity (ID) represents the unique resource. The number of bits of this identity depends on the system bandwidth and the granularity of the resource. For the system with 5

MHz bandwidth ($S$=512 subcarriers), a 2-bit resource ID are typically sufficient to represent four unique sub-bands with 14 OFDM symbols (an LTE sub-frame).

Data traffic priority represents the current traffic priority of the user [32] and should be taken into account by resource coordination. The traffic priority is a metric that is a function of the type of traffic flow (e.g., best effort, delay sensitive QoS flows), packet delay, queue length, and average rate, etc. We allocate 3 bits from the RCRM to represent eight levels of data traffic priorities.

Two bits of the RCRM are used to indicate four levels of the target SINR range of the scheduled data transmission.

It is clear that message collision occurs when two or more mobiles served by different base stations happen to send the messages with the same content, i.e., the same resource ID, the same traffic priority, and the same target SINR. When collision happens, the base station is unable to distinguish the RCRM sent by the users served by the other base stations from its own user. As a result, the base station will not attempt to do interference management for these users on this resource.

It is evident that two users from the same cell will not have the same message since the base station will not allocate the same resource to two different users. To reduce the message collision probability among users from different cells, the base station ID can be programmed into the RCRM. However, base station ID (typically 9 bits) is too long to fit into an RCRM payload. A time-varying hash function can then be used to convert the 9-bit ID into a 2-bit number. The collision probability is hence reduced.

Therefore, the resource coordination request message $m$ in an coded STS consists of 9 information bits, i.e., 2 bits of assigned radio resource ID, 3 bits of traffic priority, 2 bits of target SINR and 2 bits of hashed serving base station identity. With $S = 512$ subcarriers, the 9 bits information $m$ can be represented with $K = 1$ information symbols according to (1) and can be further encoded into a Reed-Solomon codeword with block length $N = 14$ (one LTE sub-frame). In this setup, ideally, we can see that the inequality $K \leq \left\lceil \dfrac{N}{d} \right\rceil$ holds for any *d* value up to the total number of the code words. This means, in the absence of tone erasure and detection error, up to 512 coded STS signals can be sent simultaneously without interfering with each other.

Fig. 4 shows the decoding erasure and error performance of the proposed coded STS scheme in a multi-user scenario, in which the number of simultaneously transmitting users *d* is 30, total information bits of signaling is 9 bits, the number of subcarriers in an OFDM symbol is 512, code rate of STS is (14,1) and the number of receive antennas is 1, 2 or 4. Performance in AWGN channel is plotted as a reference. SIR is defined as the ratio of received energy per sample (i.e., OFDM symbol time-domain sample) to interference (other users' data) plus noise (thermal noise) variance in time domain. An STS decoding erasure is defined as the event in which the base station fails to decode the RCRM sent from a user, while an error is an event in which the base station decodes the RCRM to a wrong but valid message. An erasure causes the base station to fail to respond to the resource coordination request whereas a decoding error causes incorrect response to the request and may result in waste of resource. In Fig. 4, the error rate is controlled below 1%. It is observed that coded STS can operate at very low SIR under multi-user (*d*=30) simultaneous coded STS transmissions. Multiple receive antennas help minimize the fading effect due to the spatial diversity resulting in performance closer to AWGN channel for the four antenna case. In the low SIR region, the up-fades from the frequency selective fading in PedB channel create more opportunities than AWGN channel for the STS tones to be detected, causing less erasures.

As earlier stated, since the transmission of coded STS is overlaid on top of other users' uplink data transmissions, the interference caused by the transmission of STS to other users is unavoidable. However, since the interference caused by STS is isolated and since STS tones are usually much stronger than regular data tones, the STS tones can be easily detected and zeroed out. In addition, isolated erasures can be effectively removed by the decoder. Minimal effect on decoding performance is expected.

Fig. 5 shows the effect of STS transmission on uplink data decoding performance. We can see that the effect on data decoding performance is less than 0.3 dB for 30 simultaneous STS users.

## V. CONCLUSIONS

In this paper, we propose a special signal, i.e., the coded STS signal, for over-the-air resource coordination and interference management in femtocell networks. The proposed coded STS scheme has many most desirable properties. First, the coded STS scheme does not require dedicated resource, i.e., coded STS overlays on traffic data and thereby does not incur system overhead. The coded STS *does* however cause interference to co-channel data transmission. However, since, unlike the conventional CDMA signals whose energy are spread over a contiguous bandwidth causing block interference to others, coded STS energy is concentrated only on one subcarrier per OFDM symbol causing isolated interference to others, therefore the effect on other users' uplink data decoding is minimal. In fact, the effect can be further minimized by erasing (zeroing-out) the subcarrier where a strong STS tone is detected. Second, the coded STS tones are orthogonal OFDM tones, therefore, users transmitting coded STS in the neighboring cells will not be blocked by the local cell users who are transmitting their own coded STS signals. Third, the coded STS is a single tone waveform with low PAPR, which is especially beneficial for coverage extension and mobile device power amplifiers. Finally, the STS is designed multi-user interference free and frequency offset tolerant.

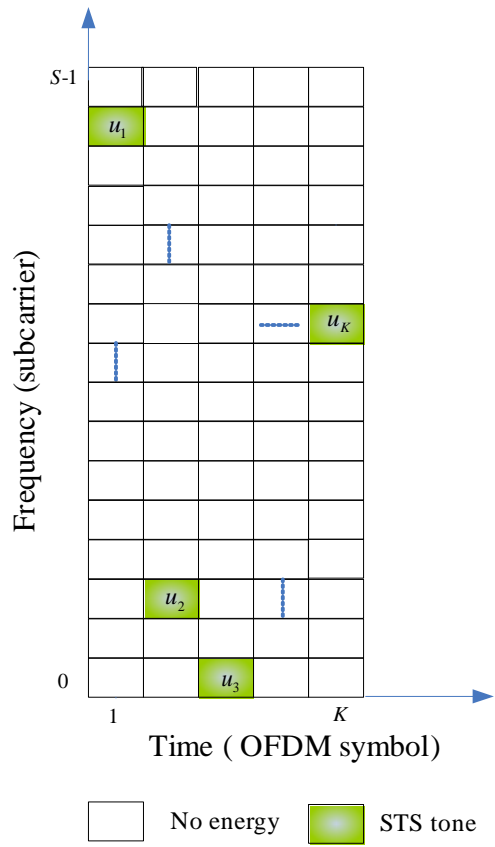

Fig. 1. Illustration of an STS signal, where $u_k$, $1 \leq k \leq K$, is the *index* of the subcarrier that the STS tone is transmitted on.

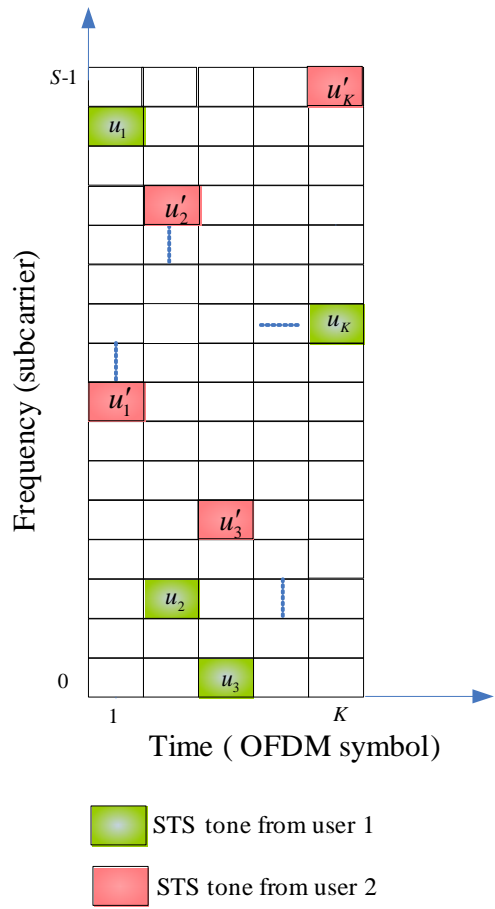

Fig. 2. Illustration of STS reception from two users.

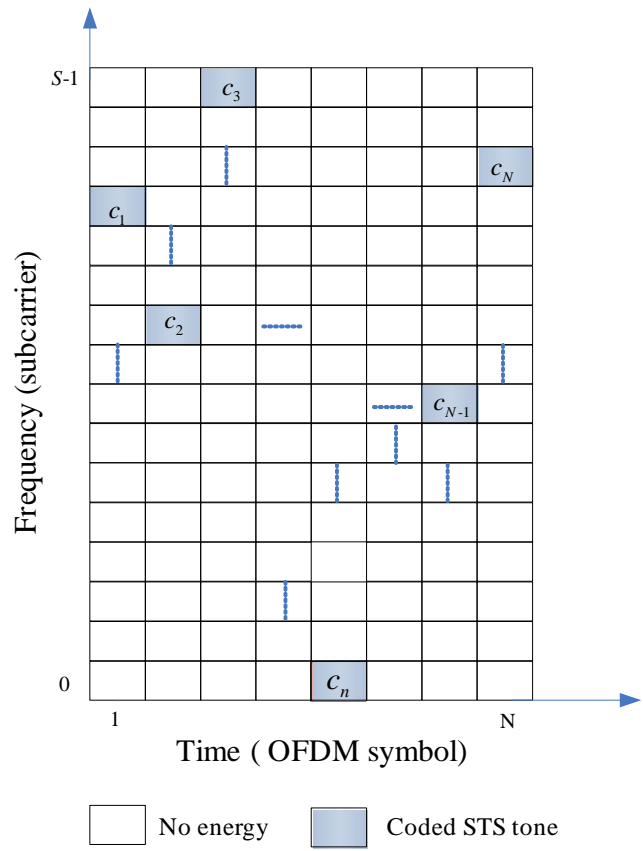

**Fig. 3.** Illustration of coded STS (note that the value of $c_n (1 \le n \le N)$ represents the *index* of the subcarrier.

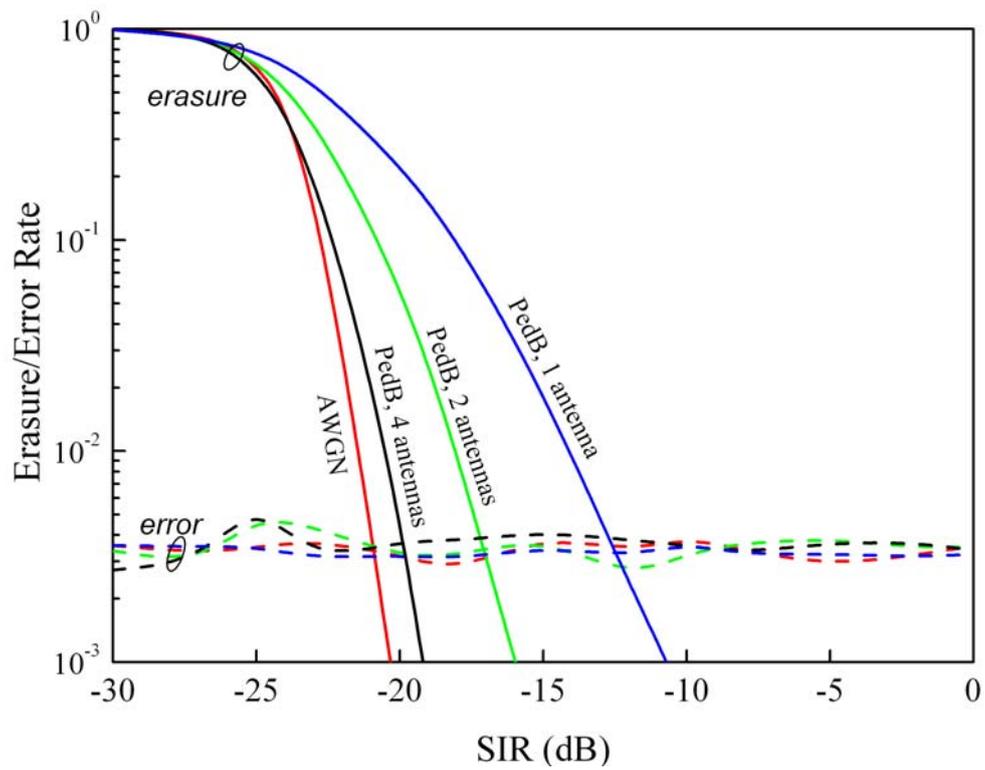

Fig. 4. The decoding erasure and error performance of the coded STS signaling in a multi-user scenario (30 users; total information bits of signaling = 9 bits; number of subcarriers in an OFDM symbol = 512, code rate of STS = $(14,1)$, fading speed = 3 km/h at 2 GHz carrier frequency; number of receive antennas =1, 2 or 4).

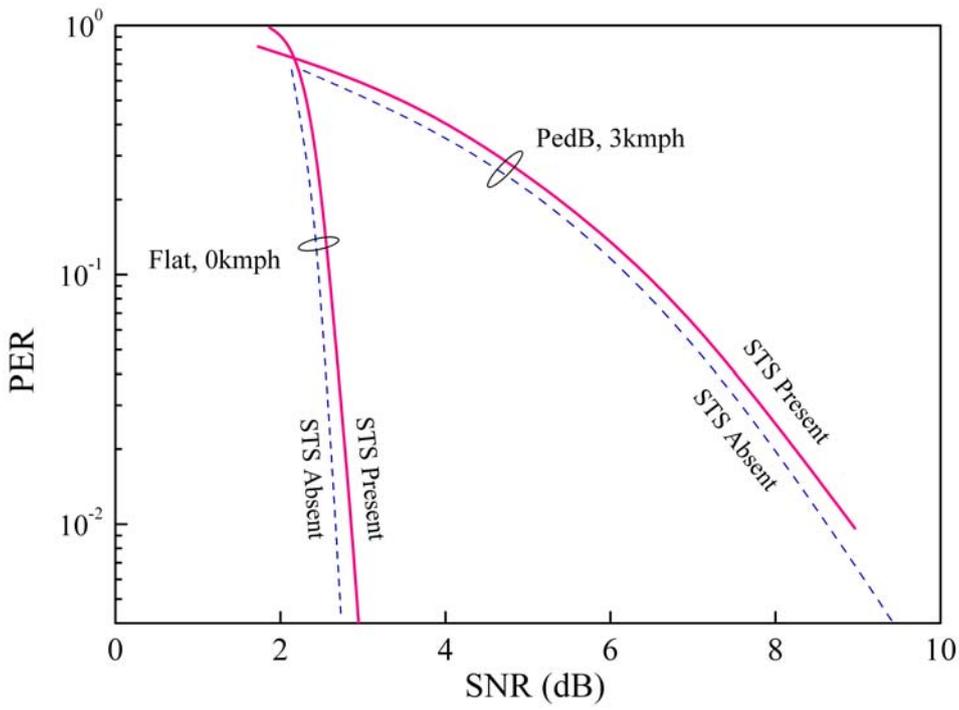

**Fig. 5. STS's effect on data decoding performance (16 QAM, 5 MHz bandwidth, 30 STS users), where SNR is defined as the receive tone SNR of uplink data per antenna.**